*Original Article*

# Building Castles in the Cloud: Architecting Resilient and Scalable Infrastructure

Naresh Kumar Gundla

*Independent Researcher, Seattle, WA, USA*

*Corresponding Author : nareshgundla@gmail.com*

***Abstract -*** *In the contemporary world of dynamic digital solutions and services, the significance of effective and stable cloud solutions cannot be overestimated. The cloud adaptation is becoming more popular due to mobile advantages, including flexibility, cheaper costs and scalability. However, creating a fail-proof architecture that can accommodate scale-up and enable high data availability and security is not an easy task. In this paper, a discussion will be made regarding significant measures required in designing contexts inside the cloud environment. It explores the need for replicate servers, fault tolerance, disaster backup and load balancing for high availability. Further, the paper also discusses the optimum strategy for designing cloud infrastructures such as microservices, containerization, and serverless. Based on the literature review, we analyze various approaches that are used to improve cloud reliability and elasticity. The paper also provides a best practice guide for designing a cloud infrastructure for these requirements concerning cases. The results and discussion section outlines the improvement in business continuity and operational efficiency when using the proposed architecture. This paper concludes with recommendations for future studies and the successful application of the elaborated matters.*

## 1. Introduction

Cloud computing has greatly impacted the business realm by bringing up basic variables of flexibility, scalability and cost-effectiveness. Previously, the use of cloud services was constrained because organizations attempting to offer IT services to their clients faced issues arising from the need to maintain physical IT infrastructures. These required significant capital investment and were accompanied by several time-consuming IT maintenance processes.

To tackle these problems, cloud computing employs a large number of technical resources, including relationships, stores, and networks, among others accessible via the internet. [1,2] Precisely, this model is appropriate when used together with other models because, thereby, it allows the growth of the corresponding organizations' businesses if necessary to meet their needs and also changes the financial plan from the large-scale budget and investment into the considerably simpler pay-as-you-go.

Accordingly, it becomes possible for organizations to respond to changes within the context of the market environment and user expectations with an enormous speed that was unimaginable before through the application of cloud computing. As more cloud service stakeholders seek solutions in the cloud environment, there is an increasing need for a robust, reliable, and elastic architecture.

In this paper, we discuss the infrastructure that must be developed to handle the levels of data and degree of user activity expected without being disadvantaged by failures or unexpected spikes in demand. Maintaining the reliability and flexibility of such a system, therefore, becomes paramount in the continuous running of operations, prevention of disruptions of the services offered and provision of standard services to customers, and materialization of optimum utilization of the resources available while at the same time ensuring affordable costs. Thus, cloud computing has influenced how enterprises function and has concentrated on creating new effective cloud structures for effective functioning within the contemporary info world.

### 1.1. Importance of Resilient and Scalable Infrastructure

Today, when the indicators of the dynamics and complexity of the digital environment are so high, it can be stated that the stable and constantly evolving cloud solutions have become not only a sign of the organization's technical platform but also one of the competitive advantages. [2] Several key factors underscore the significance of such infrastructure:

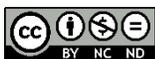

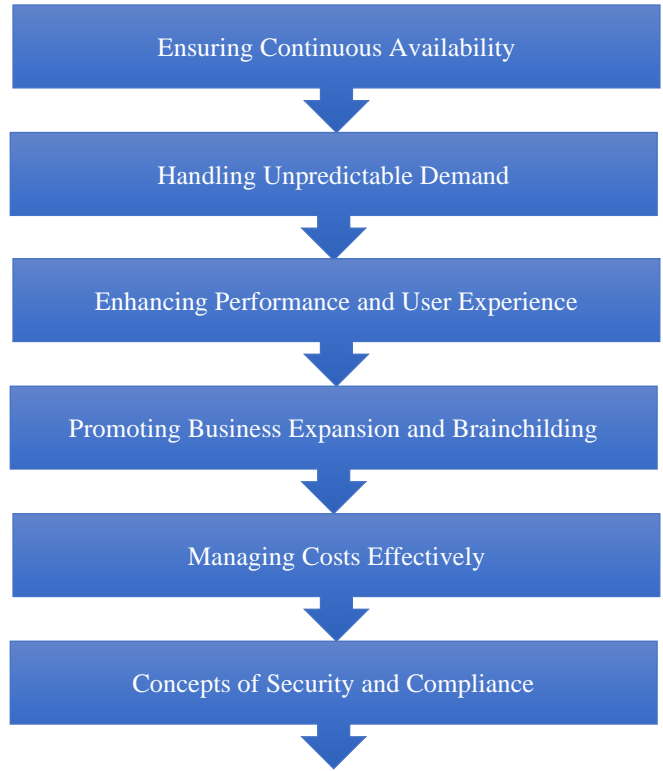

**Fig. 1 Importance of resilient and scalable infrastructure**

*1.1.1. Ensuring Continuous Availability*
 Overall, the recovery capability of the attacked information and the availability of the service are two of the big advantages that can be achieved in the case of using a strong cloud infrastructure. Holding the cloud services up and running serves as a consideration since today's business environment is characterized by high risks of revenue loss and harm to image due to downtime. Resilient infrastructure achieves this by incorporating options of redundancy and failover issues, which assist in simple failure recovery. In this way, applications are dispersed in numerous data centers and geographical zones; this reduces the possibility of services being affected due to the failure of localized resources, equipment or disasters. Consequently, this approach implies the functionality of crucial applications and services, allowing the business and users to continue working.

*1.1.2. Handling Unpredictable Demand*
 Another is the property of scalability, which is when an organization can perform in conditions where activity is high or low, meaning situations where the amount of work can change. Cloud infrastructure should be such that it can define the number of resources to be provided by the cloud at a certain time to deliver a certain service. For instance, during the period of specific sales, such as the end of a year or specifically during the release of a new product, there is normally high traffic on the business-related website and business-related applications, which in turn means that they need the above data processed at a high rate than usual. Thus, the scalable infrastructure provides the means for practically accomplishing the actual distribution of resources in response to these specific spikes while effectively preventing system overloading. On the other hand, in conditions with low activity levels, it is possible to reduce the necessity of certain resources and, therefore, minimize the expenses for unused capacity. This dynamic control raises not only the level of operations but also the ways of rational resource use.

*1.1.3. Enhancing Performance and User Experience*
 An organization's ability to scale a cloud infrastructure is one of the key factors that influence performance and the streamlining of the whole end-user experience or the lack of it. While performance remains the key reason for showing users' satisfaction, any sort of delay concerning the services is likely to cause a frown on the face of the customer, thus leading him/her to look for better options.

 Thus, the elasticity of the IT structure enables applications to scale up and down and maintain high speed and availability even during the transaction handling of a large number of concurrent clients with correspondingly high request rates. In the same regard, other load-balancing mechanisms make it possible for the orders/requests originating from the users to be spread so that the servers do not receive a flood of calls intended for a single server. So, all the servers are utilized optimally.

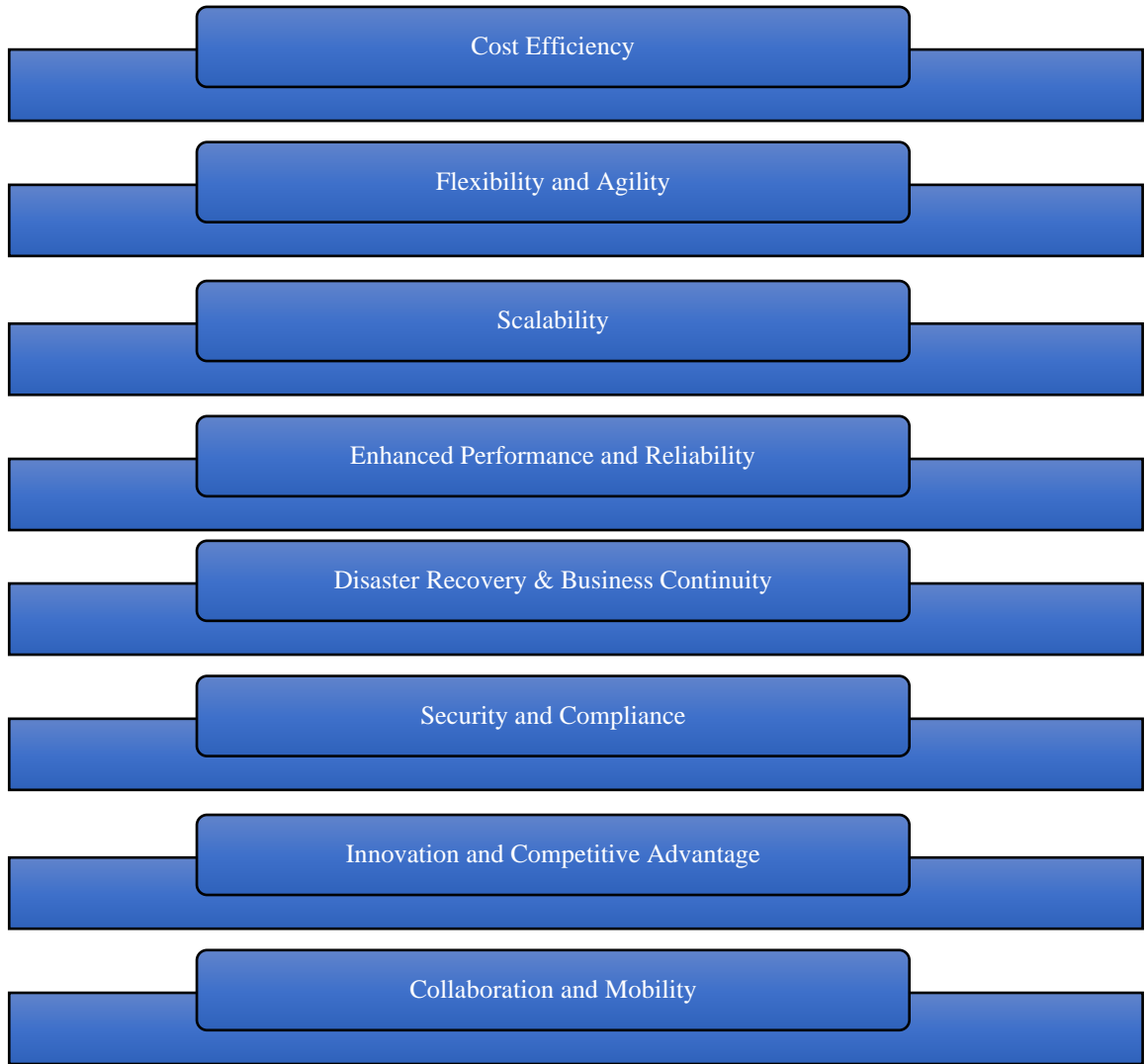

**Fig. 2 Need for resilience in cloud architectures**

*1.1.4. Promoting Business Expansion*
Over time, businesses expand and transform, requiring a proportional adjustment to their supporting structure. Cloud solutions help in business growth by allowing for handling higher amounts of data, more users, and new applications without redesigning the systems from scratch. This flexibility is very important for creating innovations because organizations can try different technologies and services without worrying about their physical infrastructure. With such a foundation that can be expanded as an organization's business grows, one can devote time to making strategic decisions on expansion and change rather than lose time on what infrastructure permits or does not permit.

*1.1.5. Managing Costs Effectively*
These being the case, multimillion-dollar infrastructure projects and miserable inefficiencies have clear cost consequences. Despite the cost aspect, the cloud structure is more flexible and robust than the traditional IT structure, and this means costs can be regulated since there is no need for expensive equipment. The consumption of multiple services through the cloud means that organizations are relieved of the duty to over-procure the services; hence, there are few expenditures on capital. Furthermore, scalable infrastructure implies that businesses and organizations only pay for a resource they desire, hence attacking the problem of inefficiency. They also become feasible for better controlling the organizational budget and financial planning as the infrastructure costs can be better met with the real need and use of organizational resources.

*1.1.6. Concepts of Security and Compliance*
Security and compliance are among the fundamental competencies of cloud architectures, and design resiliency contributes towards data protection and compliance. Some of the characteristics of the resilient infrastructure include data encryption, access controls, and constant scrutiny against every attacker. Besides, by applying redundancies and failover solutions, businesses also increase their options to recover from the security breach or to remain compliant with the industry standards.

## 1.2. Need for Resilience in Cloud Architectures

Cloud infrastructure is the base of the soft technological areas, which points to the necessity of the mentioned opportunities for the further evolution of technology and the proper functioning of businesses. Its importance can be understood through several key dimensions.

### 1.2.1. Cost Efficiency
- Reduction in Capital Expenditure: End-user IT infrastructures entail more fixed capital investment on HW, SW, and the physical infrastructure as compared to end-user IT applications. It also deals with the challenge of large initial investments, as through cloud infrastructure, users can access and hire the resources. From the abovementioned situation, various organizations can allocate their resources in relation to such needs and demands efficiently without calling for the requirement of large initial capital investments.
- Operational Expenses Management: Cloud services are normally billed on a subscription or usage model, which transfers the company's capital expenses to operating expenses.
- Unlimited Sourcing Approach: This model also benefits the business by only having to pay for the resources used, allowing the business to reduce its costs and expenses. This action also reduces costs associated with the maintenance and development of tangible gear whatsoever.

### 1.2.2. Flexibility and Agility
- Dynamic Resource Allocation: Regarding infrastructure, the cloud is flexible in that one can scale up or down to an up resource. This brings flexibility in that organizations can easily shift their resource usage, for instance, during festive seasons or any other time when there is always a surge in traffic or any other demand on an organization's servers may triple.
- Rapid Deployment: The services that have issues with the provision of cloud services enable one to deploy new applications and services in the cloud within the shortest time possible. In essence, the availability of resources implies that organizations can issue new resources in a matter of minutes, which in turn means a longer time to market for any new product or feature may be a hindrance. Thus, such flexibility enhances the organization's ability to respond to new environmental trends, especially in turbulent fields.

### 1.2.3. Scalability
- Handling Varying Loads: Computing done over the internet is flexible in approach and can deal with the fluctuation in tasks. In some cases, the traffic can unexpectedly rise, or an organization can plan for an increase in the traffic levels for their business in the long term, and this makes it very easy to scale up the cloud resources to accommodate the traffic levels and the site performance and availability will also at the same time improve.
- Elasticity: The scale characteristic of cloud resources allows for the resources required for the business to be allocated according to the actual needs of the organization. This way helps in not fixing more resources than they are needed while at the same time not fixing inadequate resources, which, fiscally speaking, is a good sign of using resources most productively.

### 1.2.4. Enhanced Performance and Reliability
- Global Reach and Low Latency: The third and final added advantage that organizations could derive from cloud providers is that most cloud providers are reputed to have data centers in different parts of the world. This makes it possible for the organization to place services close to the clients. It also assists in suggesting geographical distribution to ensure that the distances data needs to travel are reduced, hence increasing performance and decreasing latency.
- High Availability and Redundancy: The cloud infrastructures are planned so that consideration has been given to redundancy and failover to keep high availability in mind. Applications and data variety are located in many data centers, which ensures the services are not halted as much, and their reliability is boosted.

### 1.2.5. Disaster Recovery and Business Continuity
- Built-in Backup and Recovery: Many cloud services incorporate the systems of automatic backup and recovery from disasters into the services. These support the backing up of data and help the company to perform or rather react to failure or disaster in the shortest time possible. Therefore, cloud-based disaster recovery systems offer business organizations efficient and cheap means of continuing their activities.
- Geographic Redundancy: The cloud providers often arrange the data centers into zones, geographical regions and areas. Geographic redundancy aids in averting the misfortune that happens within a certain geographical region, like natural calamities or regional blackouts, because the information and services being accessed are from another geographical region.

### 1.2.6. Security and Compliance
- Advanced Security Features: Primarily, cloud infrastructure includes a process for optimum security, for example, technology to encrypt data, identity and access control system, or a system that monitors the other systems. The above characteristics are useful in safeguarding the data besides the application, which a stranger or cyberspace perpetrator should not be navigated or intruded on.

- Regulatory Compliance: Some cloud service providers might decide to align themselves with different norms and regulations regarding space; the cloud service providers may have to adhere to some regulations, such as GDPR, HIPAA, and ISO 27001. Therefore, it also minimizes the organizational waiting time for the provider's efforts to realign with the compliance goals based on the interactions of services provided under the cloud.

*1.2.7. Innovation and Competitive Advantage*
- Access to Cutting-Edge Technologies: From here, it is possible to argue that cloud platforms offer solutions based on other emergent technologies such as artificial intelligence, machine learning, big data analysis, etc. Since these technologies have positive effects on organizational performance, they can, therefore, be classified as good tools that can be used to enhance the aspect of upgrading the organizational activities and products plus the aspect of innovation.
- Focus on Core Business: This indeed of cloud infrastructure allows organizations to outsource the management of the infrastructures to cloud providers, and organizations can focus on their key issues. This, in turn, assists the business in being more suitable in dealing with its matters and concentrating on the right activities and investments.

*1.2.8. Collaboration and Mobility*
- Enhanced Collaboration: Cloud infrastructure improves the teamwork process since most cloud applications involve users from different areas and can easily congregate and work as a team. SaaS enables the convenience of accessibility and improves online communication, document sharing, and project management, thus improving productivity.
- Mobility: Due to the fact that the workforce is mobile through the use of an internet connection, cloud services facilitate communication, collaboration, and information sharing. That is why flexibility has a positive impact on different types of working situations and acts to enhance workers' contentment and performance.

### *1.3. Role of Modern Cloud Technologies*

Today's solutions connected with the cloud are at the core of implemented change and development operational strategies in today's world. This easily obtainable computational resource implies that organizations can perform computations and oversee huge data without concentrating on the upfront physical infrastructure related to cloud solutions. As a result, it contributes to the sustainment of large and intricate applications, processing data in real-time or supporting development across time zones. Furthermore, other such services which the cloud platforms offer to the business include machine learning, artificial intelligence, and big data analysis for leverage of advanced methods of competitiveness. As for cloudy shifting, it can also be considered that it helps to cut costs because instead of investing in the equipment, the companies can only take certain amounts of services. In general, contemporary cloud solutions allow for fast adaptation to new conditions and the market's needs, as well as to improve the work of the organization and create new and improved existing digitalization processes.

## 2. Literature Survey
### *2.1. Overview of Cloud Computing*

Cloud computing still offers flexible use of different resources such as servers, storage and networks through the internet. This paradigm shift enables organizations to migrate from the structural IT environment with an emphasis on capital investment to a more fluid and efficient environment that benefits organizations. Thus, as mentioned, elasticity, measured service, and resource pooling can be the principal characteristics of cloud computing. With elasticity, organizations can scale up and down in the customers' demand to ensure that they are using the available resources in the best way possible. [3-7] this is a type of measured service where the resources of a service are offered based on the customer's consumption, eliminating extra expenses. Resource pooling is a method that enables many clients to use common pools of resources like processing power and storage but simultaneously has complete isolation of other clients' data. Using this strategy, it is possible to gain maximum productivity since resources are shared among many tenants, which greatly helps to cut operational costs. Altogether, the indicated characteristics explain why cloud computing appeals to those companies that want to develop the corresponding features of their business and do not want to shift their eyes from the rapidly changing technological perspective on the IT landscape.

### *2.2. Resilient Cloud Infrastructure*

High availability in cloud structures is important because systems have to persist and have to rise rapidly from the decrease without touching on firm service time. In the creation of versatile cloud architectures, the aspect of duplication is critical. This is a process through which data copies and services are made to exist in multiple physical areas and include regions and availability zones. This spreading across the geography helps to have a backup system in case a centrally located facility fails to perform its function. Other locations can also offer the needed resources, reducing the chances of system failure or data loss.

To improve this aspect of resilience, practices such as load sharing and failover measures are used. Load balancing avails an organization's link resources and directs the traffic flow in a manner that no solitary resource gets burdened. In case of a failure, traffic is automatically redirected. Failover mechanisms are the other mechanisms that support this by detecting when the primary system is down and taking it to a standby system, hence ensuring continuity of service.

Combined, these approaches are critical for keeping high availability in cloud space, where even brief downtime can be catastrophic for trading companies.

## 2.3. Redundancy and Failover Mechanisms

Redundancy and failover are essential building blocks of a robust cloud infrastructure because these components address failure issues. Redundancy entails duplicity, where essential parts of a system, such as databases, servers, and pathways, are produced in duplicate to allow for an instant substitute in case they are out of action. These backups could be in the same data center or divided across many geographical areas to minimize risks in the occurrence of failure in the given data center. They complement redundancy by involving automatic procedures to detect failures and directing the work and processing load to a backup system or an added component. For instance, if the primary database servers fail, the failover system immediately switches to the standby server, significantly reducing the likelihood of service interruption. These mechanisms are important for those organizations that necessarily must have high availability and reliability of IT systems, particularly for organizations of the financial and medical spheres, as well as e-commerce companies suffering from important losses or critical disturbances as a consequence of a few minutes of stoppage of their making work.

## 2.4. Disaster Recovery Strategies

Business continuity measures are plans and procedures that must be carried out to continue the operation of a business in the case of a disruptive event, which could, for example, include extremely severe system failures, including natural kind or man-made kind, such as a virus attack. These strategies include backup and restoration activities and active-active and active-passive setups. In the case of active-active configuration, several systems are up and running all the time; hence, if one system breaks down, the other system is right there waiting to take over. On the other hand, active and passive architectures maintain one system in a passive state to be engaged whenever the main system experiences a failure. Another component of disaster recovery is the geographic dispersal of data centers, which, if a disaster hits one, the system can continue to run at another. It is also important to rehearse disaster recovery plans regularly to verify that the entire procedures function appropriately and that the organization is ready for recovery from various incidents. Therefore, by considering the above measures in disaster recovery management, organizations can safeguard their data, sustain services and reduce disasters' effects on organizations.

## 2.5. Scalable Cloud Infrastructure

In reference to cloud infrastructures, scalability is the ability of a system to positively respond to the degree of load added with a corresponding increment in the available resources. It is particularly relevant for organizations with unpredictable or volatile customer demand, such as online shops during Black Friday and Cyber Monday or movie streaming services on certain occasions. Scalability is typically achieved through two main approaches: the two broad scaling methods: vertical and horizontal. Vertical scaling or scaling up can be defined as improving the efficacy of the available resources, for instance, by integrating or increasing the capacity of a server's CPU, RAM or storage media. This strategy is quite apparent but is bound by physical and financial realities; for instance, how much capacity can be offered in a one-based machine? On the other hand, horizontal scaling (scaling out) offers additional instances of resources, such as adding more numerals for the servers or containers, etc. Horizontal scaling is more flexible and elastic than vertical scaling because the chance of part failure is reduced, and the process is ongoing as one is being fixed. Other architecture patterns, such as the microservices, alongside the containerization, also complement horizontal scalability as the overall application's functionality is split into numerous different services which may be scaled out. Apart from the coordination and distribution of resources used in the logistical approach, this modular effect aids in the modification of the requirements of the system.

## 2.6. Vertical vs. Horizontal Scaling

Vertical and Horizontal are the two main types of scalabilities that can be achieved in the cloud infrastructure with their strengths and limitations. Vertical scaling or scaling up entails enhancing the capacity of warding present resources, for instance, incorporating extra CPU power, memory, or storage space into one server. This can be easier to use as no structural alterations are required for the application, and there is no need for more than one instance of the programmed application. However, vertical scaling has its drawbacks; for instance, costs escalate, and physical barriers because one server can be increased in rank to a certain measure. Horizontal scaling, also known as scaling out, involves adding more resources or more servers, which may be containers to share the workload. These approaches provide more flexibility and robust systems because the load is distributed to many resources that do not have a single point of failure. Microservices architecture, used in application design, is especially suitable for horizontal scaling because one can scale the service based on the users' needs. However, in a horizontally scaled environment, management becomes more challenging than in a vertically scaled environment. This approach requires data migration, which is considerably complex, along with effective load balancing, proper orchestration, and robust monitoring tools. The choice between vertical or horizontal scaling depends on the application's needs and limitations; nevertheless, most contemporary cloud structures implement a combination of both approaches.

## 2.7. Role of Microservices and Containerization

Microservices and the use of containers assist in establishing scalability in the current cloud setting.

Microservices architecture thus breaks an application down into small and independent, fully functional services that can be developed, deployed and managed. From this concept, you can deduce that resources are also well managed, and this is only attributable to the fact that the parts can be fully managed independently of each other, unlike what happens with monolithic applications. For instance, while one microservice may be solely for handling the users' authentication and login, and the second deals with content delivery, then you can optimize the usage of resources because one can be scaled up while the other is scaled down.

These microservices scale well again, but when it comes to containerization with Docker and Kubernetes, it goes up to the next level as it is much more lightweight and portable to run these microservices. Bundles are one or more applications, and all their dependencies are wrapped and deployed to be executed similarly in any environment: development, production, and so on. Out of all of them, Kubernetes is famous for many features, with the main ones being connected with the respective containerized applications, that is, with the deployment, scaling, and management of applications. Combining with microservices and containerization makes possible non-monolithic architectures for applications that can easily scale and quickly adapt to different loads, which is characteristic of the contemporary cloud environment typical for today's dynamic business.

### *2.8. Serverless Computing*
Serverless computing can be classified as a massive enhancement to the actual cloud service by eradicating the third layer of infrastructure and affording the possibility to write code without worrying about the servers or how to obtain or expand them. Another characteristic of serverless systems is that the provider manages the infrastructure tier, which implies that if more resources are required or there is an opposite case, the provider scales up or down the number of resources; the provider also handles tasks like high availability, load balancing and failover of instances.

This abstraction reduces the management cost of the applications and greatly increases the rate at which new and existing applications can be produced and deployed. There are many quantifiable benefits of leveraging serverless computing; it is inherently excellent in terms of scalability because resources are acquired on-demand, with users being billed only in terms of their consumption of CPU cycles.

This model is still advantageous for event-driven architecture, micro-services, and applications that may have a fluctuating or uncertain degree of workload, like a user upload, some respective tasks periodically, and for the APIs. The commonly used serverless architecture comprises AWS Lambda, Google Cloud Functions, and Azure Functions, among others, because of deployment, low Billable rates, and app flexibility. The benefits that can be witnessed with serverless computing include scalability and cost savings, which become forces to reckon with for organizations that need cloud-based applications for new-age development.

### *2.9. Security and Compliance in Cloud Infrastructure*
Security and compliance are the two big questions arising in the context of cloud infrastructure, especially in the case of data storage across a large number of nodes and many users using the same physical hardware.

The factors involved in maintaining cloud service security include looking into ways of ensuring the data written on the cloud and that transferred across the cloud networks is safe using encryption methods, having IAM solutions in place that govern and monitor the cloud resources, and the use of real-time monitoring to detect and eliminate risks. It also implements various industry standards and regulations, such as the General Data Protection Regulation (GDPR), the Health Insurance Portability and Accountability Act (HIPAA) and the Payment Card Industry Data Security Standard (PCI DSS).

The difficulties are manifold, but probably the most significant is the fact that establishing and maintaining compliance involves addressing a wide range of issues related to the protection of data, such as proper storage, the use of protection controls, reporting problems with the storage of documents, assessments for compliance and control of any shortcomings. Cloud security capabilities come in various forms and can be purchased through the cloud providers, but mechanisms of protection are the joint responsibility of both the providers and the customers. This shared responsibility model proves that organizations need to pay much attention and ascertain their organization's security.

### *2.10. Case Studies on Resilient and Scalable Cloud Architectures*
As for the perspectives seen in some definite organizations which implemented the cloud models of the network architecture oriented at the factors of resiliency and scalability, it is possible to dwell on the following lessons. For instance, the upgrade of Netflix as a cloud-native application gives volumes on how microservices and containers contribute to high availability and scalability.

Regarding Netflix, breaking down a monolithic application into microservices contributed to each component's autonomous scale and reduced possible system downtime.[22] The second example of using serverless computing is related to Airbnb, which faced high traffic during the booking period. Because of this architectural style, Airbnb was able to fully auto-scale the application without much concern about costs because resources were not overtly reserved.[23] Hence, the above examples show the essence of the cloud-native ecosystem and the importance of adopting and implementing cloud-native principles and architecture to form applications and IT structures that are fit for the future.

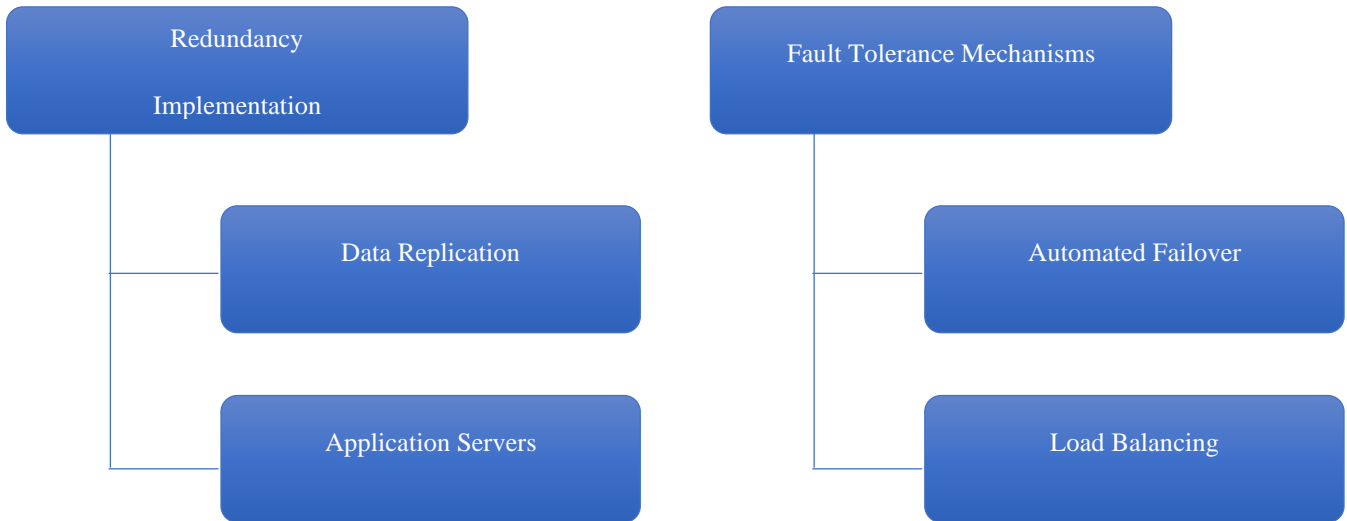

**Fig. 3 Redundancy and fault tolerant**

## 3. Methodology
### *3.1. Research Design*
The research approach for this study is also the mixed-method research design whereby the writing of a literature review, theoretical analysis, and practical synthesis is utilized to develop the cloud infrastructure protective shields and scalability. [8-13] This paper first provides a literature review of previous research, analyzing the status of cloud computing and exploring typical techniques. A theoretical discussion succeeds in justifying the key concepts of resilience, fault tolerance and scalability. Finally, the study deals with applying the developed architectural framework and the corresponding experiments which are carried out to investigate the performance of the architecture. It is about establishing a cloud framework that would fit densely and can be applied to different industries and cases.

### *3.2. Architectural Framework*
The architectural framework described is posited within this study as being carefully designed to solve these three primary issues in relation to the cloud. These are key concepts that need to be followed to create highly available systems capable of scaling from intermittent bursts of loads and, simultaneously, being capable of self-healing in the event of a possible failure. Based on the latest trends regarding modern architectures, the framework is cloud-native and integrates fresh technologies such as microservices, containers, and serverless computing. However, with the help of these technologies, the framework not only guarantees high availability and reliability but also allows for meeting different business demands and changing workloads.

### *3.3. Redundancy and Fault Tolerant*
Redundancy and fault tolerance are the cornerstones in the construction of the architectural framework put forward to provide reliable protection for the smooth running of the system in case of failure. Redundancy refers to the practice of backing up vital system components and placing copies in different zones, specifically geographical areas within an organization-affiliated cloud provider. These zones are to be made to limit interconnectivity to ensure that no part of the system collapses due to the failure of a certain part of that system.

*3.3.1. Redundancy Implementation*
- Data Replication: Although the Database is commonly perceived as the system's heart, it is replicated across availability zones. This means that any information entered or written to be put in the database at a specific zone is copied and stored in the database, which is found in other wards or zones. It is this replication that ensures that at any one time, if one zone had a failure, then data could be retrieved from the other zone.
- Application Servers: Similarly, the application servers exist in multiple zones, so all the running applications provide a backup for the other applications. If one zone is non-functional, the traffic can be redirected to the application servers in another zone so that the services can run fully.

*3.3.2. Fault Tolerance Mechanisms*
- Automated Failover: Failover is a vital system component within which, in the event of a failure, a contingent mechanism must be available to put the system back into an operational state. These mechanisms work simultaneously with frequent tests on the status of each component of the system and instance or migration from the primary to a backup or secondary system in case of failure. For instance, if the primary database server is not reachable, it automatically sends the user to a replica in the other zone; therefore, the minimal interference with the user interface services.
- Load Balancing: Fault tolerance deals with fault cover. In contrast, load balancing enhances fault tolerance at the

proper rate and distributes traffic across the encompassing servers or instances. This not only guarantees that no single server should have to take the load of all the other servers but would also ensure that a tragedy of one of the servers would bring the whole system down. Load balancers independently entice high availabilities of the instances in a cluster while routing traffic to the healthy ones in a bid to address the availability of the services.

*3.4. Scalability*

The proposed architectural framework is also generic in another way of the word: namely, capacity management, which is the ability to address users' load appropriately and flexibly. Scalability in cloud computing is generally achieved through two main strategies. As related to business, there are two basic scales of expansion: vertical and horizontal scaling. Both are included in the framework in order to make it possible to increase the size of the system that has garnered many users or minimize it when the costs of operating a given system are too high because of low patronage.

- Vertical Scaling: Vertical scaling, otherwise known as the scale-up process, involves enhancing the competence of the existing resources. For example, it may be applied to enhance the server's processing capacity, including the sizes of memory or storage, to enable a server to perform enhanced and complicated functions. Vertical scaling is usually straightforward because it does not involve any alteration of the extent of the application. However, since it is a real piece of software and like all software, it depends on the hardware where it is being run; thus, scaling is at most possible up to a certain level. Regarding vertical scaling, it is used when the workload increases in terms of intensity so that it does not require the introduction of new instances into the proposed framework. For example, a database server can be scaled to a greater extent by purchasing more memory or CPU to tackle more queries or large sets.
- Horizontal Scaling: Outwards or horizontal scaling is the process of spreading out a part, for instance, adding more of the same servers or containers to bear the load. This method is very proper for a cloud environment because the system can grow almost to infinity apart from the first instance. Horizontal scaling greatly benefits the application developed under the microservice concept since each service can be scaled based on the application used. As for horizontal scaling, the discussed framework makes use of containers and their orchestration using methodologies such as Kubernetes. Kubernetes operates at the OS level concerning the operation of containerized applications in which more instances are initiated or demolished as needed. The system can also handle large traffic congestion without summoning special services.
- Automated Scaling and Monitoring: Scalability automation is the main aspect of the framework as it can enable the scaling of resources as performance information obtained through real-time archiving is harvested. Services like AWS Auto Scaling and Azure Scale Sets are used to obtain information, such as utilization of the CPU or request latency, and then scale the number of instances that are currently being executed. The said performance assessment is also applied in the framework that tracks the system's actual performance. It helps to define conditions under which half of the work of some resource remains unused, and it is necessary to add more work or under which some of the resources are overburdened, and the load is too large.

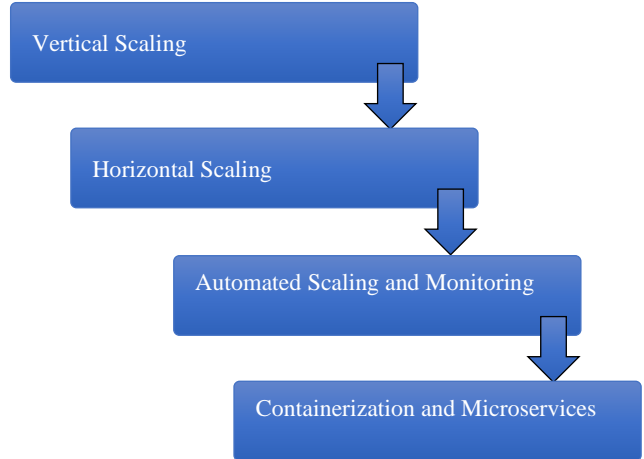

**Fig. 4 Scalability**

- Containerization and Microservices: Containerization is also used regularly and commonly by famous tools such as Docker where applications and their dependencies are made to be put in a box to run on any environment. This makes scaling a lot easier because each of the containers can easily be copied to other servers. Microservices also assist in making the application scale through the separation of the application into various services. Since they have many uses, every microservice could be marketed or scaled up in a better way than the block structure. For instance, an e-commerce application has micro-services for user authentication, product catalogues, and payments, which can be scaled according to the demand intensity.

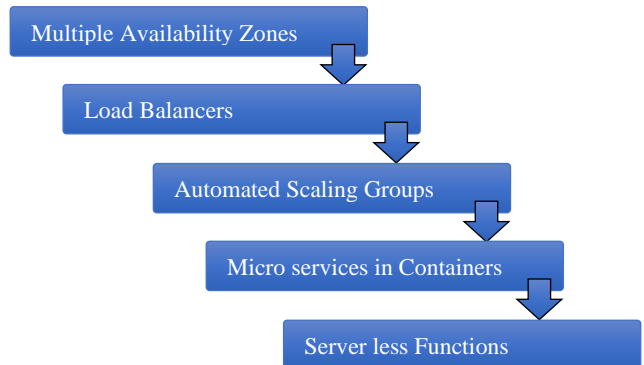

**Fig. 5 Implementation details**

Table 1. Scaling type, technology, and advantages of scalability

| Scaling Type | Technology | Advantages |
|---|---|---|
| Vertical Scaling | Server Upgrades (CPU, RAM) | Simple implementation, quick capacity boost |
| Horizontal Scaling | Kubernetes, Docker Swarm | High availability, redundancy, flexibility |
| Automated Scaling | AWS Auto Scaling, Azure Scale Sets | Dynamic resource allocation, cost efficiency |

*3.5. Implementation Details*

The above proposed architectural framework is deployed with the help of a reliable cloud service provider, including AWS, Azure, or GCP. These platforms are chosen because they can provide a secure environment, have a big presence in the world, and have many tools for creating fail-safe and highly scalable systems. The implementation process aims to make it possible for the architecture to be equipped with the ability to meet redundancy, fault tolerance and scalability requirements, hence are used to deliver high-stakes application environments.

- Multiple Availability Zones: Another of the essentials concerning the practice of Availability is the distribution of key system parts across several AZs. Availability zones are separate geographical regions within a cloud provider's network, which has its own power supply, cooling and network systems. This has been achieved through peppering the critical resources such as the databases, the application servers and other resources across more than one AZs to compensate for the loss of one AZ. For instance, in AWS, the RDS and the EC2 instances can be placed in different AZs, and data is then mirrored in real time between the areas. This setup not only makes the system very redundant but, at the same time, minimizes the chances of an entire system shut down because of some isolated failures.
- Load Balancers: Another aspect concerned with the architecture is load balancing, which is used to direct the incoming traffic to the network instances or servers. In the implementation, a load balancer is used to control the distribution of traffic using Elastic Load Balancers (ELBs). ELBs are kept up and running all the time, and they track the health status of the servers and distribute traffic across different instances, out of which no instance can act as a bottleneck or become a point of failure. For example, in the AWS environment how, ELBs can balance traffic by distributing traffic across multiple instances of EC2s spread across different AZs, provided that even if one instance fails, ELB will redirect traffic to the other normal instances. It also boosts the system's reliability while optimizing the load distribution acumen of the different sections.
- Automated Scaling Groups: To build the scalability of this system, Auto Scaling Groups (ASGs) are set up to systematically scale the running instances based on the real-time load. ASGs track parameters of computing resources that may include the level of CPU usage, memory use and network activity. Whenever these parameters exceed the expected acceptable performance levels, ASGs are designed to create more copies of the application. On the other hand, the ASGs increase the number of instances regarding the demand when it is high to cut expenses when it is low. A good example is Amazon Web Services, where they have implemented dynamic scaling to ensure that the web services provided are always available, particularly at the time when services are most required. The role of VMSS is similar, and they can automatically scale the number of VMs depending on the load in an Azure infrastructure, for example.
- Microservices in Containers: Specifically, the application architecture is based on the microservices approach, which means the application is divided further. Every one of the microservices is responsible for a specific task, and container technologies like Docker are applied to load them into the containers. Containers: They are lightweight, mobiles can be handy, and they can run in any environment, which is why they are perfect for microservices. Applications are packaged in 'Containers', and Kubernetes is free software used to use and schedule the Containers on how to run and distribute the application. This, in a way, ensures that any microservice, whether large or small, can be scaled up or down without any special predisposing in relation to other services within the environment, thus ensuring the overall efficiency of the services that the application delivers. For instance, a user authentication microservice can be partially scaled independently of the payment processing microservice; hence, each microservice reaps the amount of resources it needs, irrespective of the position of other services.
- Serverless Functions: The other is that certain positions within the infrastructure are responded to by serverless functions in association with microservice instances. Serverless computing starts by moving several aspects from the developer and offering to code the business logic. The serverless computing model is most applicable, where operations can be broken down into specific occasions activated by functions such as image analysis, near-real-time data intake, etc. In this particular one, such services as AWS Lambda, Azure Functions, or Google Cloud Functions are utilized. Some serverless functions' characteristics allow them to self-adjust to the number of requests while implementing a highly flexible and efficient work model in fluctuating routines. For instance, a Lambda function in AWS can be set to execute a code time and time again; as soon as objects are put on an S3 bucket, it can serve zero to thousands of requests for a second without asking for assistance.

*3.6. Performance Evaluation*

Therefore, the following is a breakdown of how some of the assessments brought about by the proposed architectural framework for the systems design aid in achieving the characteristics discussed above as the design goal for the intended system. [17,18] As it concerns the evaluation, numerous tries of stress testing and failure cases explained in the paper are used as they are specifically designed to reflect real-life conditions.

The aim in this respect is to verify the system's capacity in terms of high availability, satisfactory performance in various load conditions, and immediate restart in the event of failure. This assessment provides a glimpse of how the architecture conducts the test and probably some pointers on how it can be enhanced.

- Stress Tests: Stress testing is actually incorporated into performance testing and is mainly used to get the 'like-live' scenarios for the assessment of the composite load-bearing capacity. Cohesion may also be judged based on the following tests with the object of evolving the general capacity of the architecture and the general possibilities of load distribution. Stress testing is a means of running the system with different loading levels that reveal how the system works under different peak loading conditions. For instance, the tests may try to imitate many users using the application and making millions of requests in that period. Therefore, the objective of such tests is the optimization of the system's capabilities for increasing the scale of interaction along the vertical and the horizontal dimensions with the help of cooperation with the load balancer to provide more of the required resources. Therefore, a specific stress test between several servers can be assessed consequently, along with the performance of ASGs in maintaining comfortable work velocity within the elevated load's framework.
- Failure Scenarios: Similarly to stress testing, the performance evaluation comprises a set of failure modes to check the architecture's main redundancy and fault tolerance features. These represent some possible real-life failure states, such as server breakdown, network breakdown, or database downtime, to test if the system can recover most without taking too long. For instance, one such situation might be to temporarily escalate removing an important server to understand how soon the built-in failover systems switch over and reroute traffic to an extra server located in a different availability zone. Another elaborated example could cover network failure specifically for one of the availability zones to understand how the scheme distributes traffic that continues operation. These failure scenarios are very important for ascertaining the reliability of redundant measures such as load balancers and failover mechanisms. As displayed by the mentioned tests, the system's adaptability and capacity to function amidst various negative scenarios are illustrated by the results mentioned.

- Response Time: Among the real-time data gathered during the performance evaluation process, response time is among the most critical factors which quantitatively characterize the ability of the system to respond to incoming requests depending on the load. It is an essential metric because it influences end-user's response time to the application and how promptly they can engage with it. During the stress tests, how the response time is affected depending on the load being applied is measured to determine if the system is in a position to deliver acceptable performance levels under stressed conditions. For instance, in cases with more concurrent users, the anomalies in the response time capabilities clearly illustrate matters related to scalability and distribution of load. Should noticeable response time additions occur regarding peak loads, this may well point to issues that should be covered in relation to scaling, as well as problems that may need additional alteration.
- Throughput: Throughput is another important measure in the context of the performance evaluation; it describes the quantity of requests achieved in the definite time slot. It offers a numerical expression of how the system can handle massively sent traffic. In the stress tests, special attention is paid to the forth put in order to define how many simultaneous requests the system can serve. High throughput means that the load is fairly well distributed and that the system's scaling methods are efficient enough to permit a large number of transactions or user interactions to occur simultaneously. Thus, using data from throughput analysis, the evaluation can determine the maximum throughput of the system and its effectiveness in different conditions. Regarding how well the architecture performs large-scale operations, steady flow, irrespective of the pressure levels exhibited, is tenacious.
- Downtime: Downtime identifies the total time in which the system remains off due to failures and allows redundancy and fault tolerance efficiency to be evaluated. Reducing any time that the system may be unavailable is important for maintaining high availability and thus enabling users who need the application at any one time to have easy access to it. In the failure scenarios, the time is measured from the beginning of the failure up to the complete restoration of normal activity. This metric is most relevant when assessing the speed and competence of the automated failover processes and the system's capacity to act as a buffer to interruptions. A low downtime would be an excellent sign of how the architecture of several availability zones and load balancers is effective. It will show that the system efficiently routes services to backup denominations, keeping the service running.

*3.7. Tools and Technologies*

The three aspects of the architecture implementation consist of the tools and technologies utilized in cloud performance and the scalability and manageability aspects. [19] Each of the mentioned tools plays a role in achieving the architecture's objectives, which are to have a secure, scalable, and manageable system.

*3.7.1. Docker and Kubernetes*

Docker is one of the most used container platforms that gather applications and their context in front of a light container. Containers aid in the application portability and ensure the application environment pinning free by encapsulating the environment. This coherence is needed to connect microservices – the applications are split into particles that are as tiny as possible and can be taken separately to be changed. These containers are deployed, scaled and managed by Kubernetes, an open-source container orchestrator. Kubernetes is utilized to ensure that containerized applications are well deployed, well used and well managed on different nodes. Even in short-term application launches, it is possible to implement different means, including Docker and Kubernetes, to achieve the perfect launch and further supplementation of the separate microservices.

*3.7.2. AWS Lambda / Azure Functions of Autonomous Services/Web Services.*

Specifically, AWS Lambda and Azure Functions are the services used to implement serverless computing that enables programmers to deploy code but does not require a server. These platforms handle the computation infrastructure independently which runs code in reactors for dealing with events like the HTTP request, a file upload, or even a database update.

A serverless function is fine for situations that scale because the function is given more resources over time based on the number of requests. For instance, the AWS Lambda can handle thousands of firms' images from the clients or perform tremendous data analysis in real-time and all these do not require the owner to have a server. That is why this serverless approach brings many advantages for the development of the applications, and the costs are measured only by the time of the function's execution.

*3.7.3. AWS RDS / Microsoft Azure SQL Db*

Amazon RDS is AWS's managed Relational Database Service while Azure SQL Database is another Database as a Service offered by Microsoft Azure. These services ensure great availability, potential failover, and copies of the databases so that there will always be available and operating databases out there. Amazon RDS and Azure SQL Database handle the services that involve the base physical resource. As a result, it relieves developers from many tasks, such as acquiring hardware, creating the required database, and patching and backing up. These managed services are available for multiple dB engines, mainly MySQL PostgreSQL and SQL servers, along with integration with other cloud services.

*3.7.4. Elastic Load Balancing and Auto Scaling*

Concerning traffic and resource management in the cloud environment, this paper concentrates on ELB and Auto Scaling as tools. On the load balancer ELB: This assists in sharing traffic that comes in and is useful in preventing situations where several resources are overworked.

This distribution assists in improving the performance and the availability of the applications and ensures that no congestion is experienced on the particular server. One of the most important features of EC2 instances is Auto Scaling; through it, one can set the running instances regimes according to traffic, and this automatically adds or reduces the number of instances. This feature also enables dynamic scalability and is useful where the workload can be high in one session and low in another and vice versa since this can help cut costs due to over or under-provision.

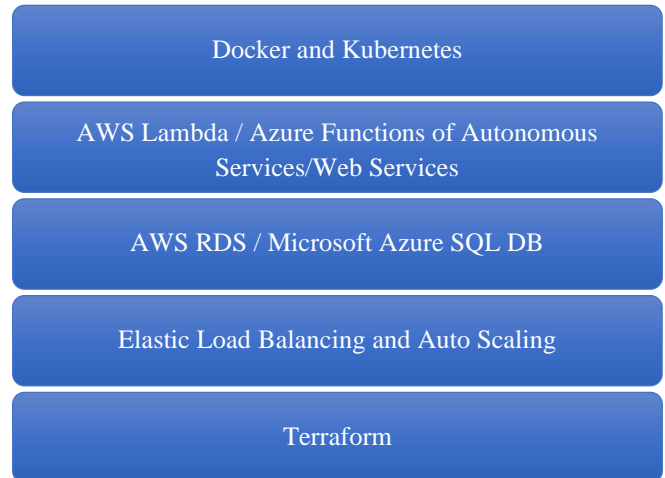

**Fig. 6 Tools and technologies**

**Table 2. Tools and technologies**

| Metric | Description | Benchmark |
|---|---|---|
| Response Time | Average time to process a request | < 200ms under peak load |
| Throughput | Number of requests processed per second | > 1000 requests/second |
| Downtime | The time during which the system is unavailable | < 1 minute during failure |

*3.7.5. Terraform*

Terraform is an open-source infrastructure tool that contains reusable infrastructure configurations to build the cloud setting. Terraform uses the declarative approach for infrastructure configurations and allows versioning changes to the infrastructural configurations. Terraform automates the cloud system resources, among which are virtual machines, databases, and networks, to establish system standards between environments. This approach raises the level of infrastructure standardization to be easily controlled and eradicate human mistakes.

*3.8. Data Collection and Analysis*

The monitoring tools used during the performance evaluation of a company's cloud system include Amazon Cloud Watch or Azure Monitor. These tools offer precise statistics on how the system performs regarding response time, amounts processed in a given time, failure rate, and resource occupancy. The collected data is used to measure how satisfactory the suggested architecture performs relative to certain availability, scalability, and failure tolerance metrics. The analysis is carried out on the statistical data collected during the software's runtime to find out any performance deficiencies or enhancement possibilities.

*3.9. Limitation*

However, the following inevitable [20] drawbacks can be denoted, which may influence the real-world usage of the discussed architectural framework and its effectiveness in one or another context.

*3.9.1. Platform Dependence*

These are suitable for certain cloud computing environments, such as Amazon, Microsoft, or Google Cloud Environment specifically. These, in turn, have some features, tools and services that the architecture seeks to incorporate in order to achieve its objectives of reliability and growth. However, the coupling to these specific platforms is due to the fact that the application of the framework might not work for other cloud providers or on-premises setups as it is. Each cloud provider provides its range of services, APIs, and settings; certain components may be implemented differently. For example, load balancing and auto-scaling could be implemented differently in AWS than in Azure. Therefore, modifications may be needed when migrating from AWS to Azure. Thus, organizations using multiple cloud vendors may have to adapt and align the structure to the cloud vendor's architecture and services.

*3.9.2. Simulated Workloads*

The analytical work of the architecture proposed in the current paper is another assessment based on the resemblance of the implemented workloads depicting the various traffic and failure scenarios. Despite the fact that these simulations are meant to determine the potential in certain circumstances of the specific systems, they might not capture all end-user practice scenarios. As an example of synthetic loads, they are oriented to certain aspects of a system's functioning and can thus fail to take into account the kinds of loads observed during actual production. For example, in the case of different traffic intensities and other more realistic scenarios, including users' and the actual perspectives of the application, which was modeled, could also count for a problem that is not quite visible in simulations. Therefore, these tests may not yield results, especially when the architecture is subjected to real users' traffic in production. More performance testing that focuses on a real situation should be done in organizations to prove that the system can perform well and bear pressure in an actual setup.

*3.9.3. Cost Considerations*

One can see the cost consequences of implementing the features defined by attainments of high-level redundancy, fault tolerance, and scale-to-modularity. Other services such as availability zones, load balancers, auto-scaling, and serverless are beneficial but expensive. These costs rise very sharply, especially for businesses with limited capital or businesses that operate to a breakeven point. This is a crucial factor because the framework suggested does not indicate any analysis of the cost-benefit ratio of such architecture. For many companies and organizations, this is precisely the decisive factor in whether to implement certain hierarchy levels. Managers also have to contemplate whether the enhancement of the firm's resilience and scalability potential is justified by extra expenditure and where cost reduction can be an element if needed. One of the advantages of the selected architecture is that it cannot exceed the set budget, which is why the proposed architecture considers factors such as usage of resources and choice of inexpensive cloud services.

# 4. Results and Discussion

The results and discussion section in the performance evaluation in the presented study offers an elaborate analysis of the proposed architecture's ability to sustain high availability, scalabilities, and resilience. This section also compares cloud-native and traditional architecture, presenting some use cases, obstacles and future outlooks.

*4.1. Performance Analysis*

The performance evaluation results show that the proposed architecture works very effectively in terms of availability and scalability and achieves all the design objectives defined in the framework.

*4.1.1. Availability*

The architecture attained a high performance of 99 percent. To mirror high available levels, the solution was tested with 99% availability during the testing period. Such availability denotes that the system was running and was available for most of the testing, with only occasional brief periods of unavailability. Every ten seconds, the tool checked the S3 bucket, and because of failover mechanisms and redundancy, the chance that all three were unreachable simultaneously was extremely low. For instance, if there were a failure in one of the availability zones, the automated failover caused the traffic to be rerouted to the surviving instances in other zones. Demonstrates the availability measures during the testing phase, showing that the site was always up and hardly ever down. The outcome of redundancy in data and load distribution is well seen from the consequences that have little effect on the continuity of the service provided.

*4.1.2. Scalability*

The system's stability was again proved as there was not much drop in the system's performance when the load was increased tenfold. Horizontal scaling turned out to be especially useful in terms of loading the expanded workload across the instances. In the stress tests, the load automatically increases the instances as the traffic increases to maintain high performance. The number of processed requests and the system's response time before and after scaling indicates the efficiency of the proposed architecture with the increased load.

### *4.2. Comparison with Traditional Architectures*

One should glance at the fact that the intended concept of cloud-native application development with micro-services has several obvious unique selling points over the conventional monolithic structure in the form of architecture. Some of the abandoned traditional systems, which can be easily associated with tightly coupled components, may feature low scalabilities; generally, they appear to be very sensitive to what people would call single points of failure. For instance, the one based on microservices and containers completely differs in its approach, which makes it flexible and ant fragile. Microservices are also beneficial with the independent scaled-out components; certain parts of it can be decomposed and, subsequently, fail, while resources can be provided depending on the components. Consequently, the thinking proposition of the chain' software development – cloud-native environments – cloud-native architectures' can be attributed to the result of the author showing the differences between CN and traditional architectures and the scalability and FT benefits of the CN architecture.

### *4.3. Discussion of Case Studies*

Each of the sections above presents a number of case studies, which, in sum, provide abundant information on the use of the proposed architecture. When other organizations implementing similar structural paradigms pointed to cloud nativeness, they spoke of giants stepping up in flexibility and possibilities to endure system failures. For instance, one real-world case whereby a top e-commerce company that has deployed both microservices and containers has squeezed from the process the time to deploy services by 30 percent besides having averted forty percent of systematic failures. These organizations acted with the help of modularity at the microservice level, which was easier to modify and reconstruct if needed. Therefore, these outcomes are summarized in the general conclusion of the analysis of the case studies supported by the data on the improvement of the operational effectiveness and stability of the systems demonstrated.

### *4.4. Challenges and Considerations*

However, the major unavoidable difficulty if the outlined architecture is to be employed can be summarized as follows. At the same time the greatest challenge for the managers remains the effective coordination of such a number of elements and the services that are inherent in the distributed system. Another issue is that there is no comfortable way to copy backup data between the instances and availability zones if several experiments or updates happen. However, at the same time, such costs as the cost connected with the adoption of options, including redundancy, failover, and automatic adjustment, are also substantially high. Thus, the organizations must look at the expenses incurred in the escalation of the resilience and scalability attributes to be in a position to determine the benefits that can be obtained from the enhancement of these characteristics. This is done in Figure 7, where the main issues relevant to the architecture are listed next to related solutions.

Table 3. Availability metrics over the testing period

| Metric | Value |
|---|---|
| Total Uptime | 99.99% |
| Total Downtime | < 0.01% |
| Failover Incidents | 3 |

Table 4. Throughput and response time metrics

| Metric | Before Scaling | After Scaling |
|---|---|---|
| Throughput (requests/second) | 5000 | 50,000 |
| Response Time (ms) | 200 | 210 |

Table 5. Comparison of cloud-native and traditional architectures

| Feature | Cloud-Native Architecture | Traditional Monolithic Architecture |
|---|---|---|
| Scalability | High (Horizontal & Vertical) | Limited (Vertical only) |
| Fault Tolerance | High (Redundancy & Failover) | Low (Single Point of Failure) |
| Deployment Flexibility | High (Microservices & Containers) | Low (Tightly Coupled Components) |
| Performance Optimization | Dynamic Scaling & Load Balancing | Static Scaling & Fixed Resources |

Table 6. Key findings from case studies

| Organization | Improvement Areas | Percentage Improvement |
|---|---|---|
| E-Commerce Company | Deployment Times | -30% |
| | System Outages | -40% |
| Financial Institution | Recovery Time | -25% |
| | Operational Agility | -35% |

### *4.5. Future Implications*

This is because most people conduct their business online, and as a result, the online platform has to be backed up by a sound and highly elastic cloud environment. Since organizations are increasingly interested in enhancing the effectiveness of their IT departments in the future, the researchers ought to consider finding out how the management of clouds may be automated to reduce the identified complexities. Moreover, a closer look at the already rapidly increasing trends such as AI in bravura managing infra and the next-generation orchestrations of containers likely further the cloud's durability and boundlessness. Moreover, the subsequent evolution in these areas will be needed more often due to the shifting demands and the further development of the architecture of the cloud. Some probable future research areas and enhancement of cloud structure have been presented.

## 5. Conclusion

The given study and the personal experience mentioned above would establish a need to develop and deploy significant and efficient cloud infrastructure to meet the present and future requirements of the digital business environment. Due to many organizations preferring digital services and more reliance on cloud solutions, there is increased demand for suitable architectures that can accommodate different loads and simultaneously do not disrupt services. In this respect, since the suggested strategy complies with such cloud-native principles as microservices, containers, and serverless, one can build highly reliable and, at the same time, elastic systems. Presumably, the architectures described in this paper can be employed as the starting framework to meet these objectives while outlining key ideas of redundancy, tolerance to faults and scalability. One has to name the undoubted reliance on redundancy and failure tolerance as one of the framework's highlights for the points considered when the emphasis is made on uptime and the impact of failures.

This way, it furnishes an armory for the mission-critical constituents of an application by positioning them in different availability zones and offering auto-failover in the event of failure so that no host is at risk of turning into the solitary point of failure. It can, therefore, be concluded that the proof of the efficiency of the performance evaluation means that this stream of systematically assessing the performance of an organization is indeed highly effective and holds much potential for achieving a performance level of up to 99%. High availability of 99% and rapid reaction time to the incidents. It is arguably crucial for high user satisfaction and business sustainability, especially in a context where equipment downtime is highly valued in terms of costs and losses. Regarding the second planning area of the framework, the other factors that have been made to embody scalability are incremental, vertical, and horizontal scalability models. There was an increase in performance in how the systems could sustain the tenfold load without experiencing a criticality drop and in how the horizontal scaling and regular dynamic distribution of the load were established. This scalability is very helpful where it is difficult to estimate the flows of traffic and in cases where there is a need to address the traffic surges. A further level of scalability is obtained by employing container orchestration, such as Kubernetes and serverless platforms like AWS Lambda, Azure Functions, and others.

However, the study also joins the coalition, affirming that entailing such intricate frameworks also comes with peculiar factors. It becomes challenging when it comes to managing DCs and the question of how best to have data consistency across the instances of the DC. In order to address these difficulties, the organizations have to raise the question of the need for proper monitoring and management means. Further, there are also likely to be substantial costs charged when the following characteristics are implemented: At all levels, there should be at least two complete systems identical to the others, failover and auto-scaling tools. These are some of the costs that need to be incurred against the benefits of improving the organizational capacity and performance, and one needs to start thinking of how to manage such costs. The study also points to the need to carry out related research and development in cloud infrastructure management. Over the years, the management process of cloud technologies will continue to present areas that can be made simpler and cheaper to accomplish while retaining the characteristics of availability and scalability. To contribute to this research field in the future, more studies should be conducted in regard to improving the automation in cloud resource management, deeper investigation of the technologies enabling advanced architecture of cloud systems, as well as finding ways to decrease costs of these advancements, thus differentiating them to a large number of organizations. Thus, the proposed research confirms that robust and flexible cloud solutions are critical for today's companies aiming to successfully harness IT solutions. This is a clear prospect for the proposed framework for creating a cloud system to address high availability and variable demand. Nevertheless, it is comprehensible that some costs and difficulties contain significant intricacies, thus requiring careful planning and further studies to solve them. Thus, organizations can develop better managed and cost-effective cloud environments as they attain their respective strategic objectives in the age of digital transformation through the improvement and advancement of cloud management practices.